\documentclass[twocolumn,aps,prl,amsmath,amssymb,floatfix]{revtex4-2}
\usepackage{graphicx}
\usepackage{dcolumn}
\usepackage{bm}
\usepackage{xcolor}
\usepackage[normalem]{ulem}

\begin{document}

\title{The lifetime of charged dust in the atmosphere}

\author{Joshua M{\'e}ndez Harper$^1$}
\author{Dana Harvey$^2$}
\author{Tianshu Huang$^2$} 
\author{Jake McGrath III$^2$}
\author{David Meer$^2$}
\author{Justin C. Burton$^2$}

\affiliation{1. Department of Earth Sciences, University of Oregon, 1255 E 13th Ave., Eugene OR, 97403, USA}
\affiliation{2. Department of Physics, Emory University, 400 Dowman Dr., Atlanta GA, 30322, USA}

\date{\today}

\begin{abstract}
Windblown dust plays critical roles in numerous geophysical and biological systems, yet current models fail to explain the transport of coarse-mode particles (>5 $\mu$m) to great distances from their sources. For particles larger than a few microns, electrostatic effects have been invoked to account for longer-than-predicted atmospheric residence times. Although much effort has focused on elucidating the charging processes, comparatively little effort has been expended understanding the stability of charge on particles once electrified. Overall, electrostatic-driven transport requires that charge remain present on particles for days to weeks. Here, we present a set of experiments designed to explore the longevity of electrostatic charge on levitated airborne particles after a single charging event. Using an acoustic levitator, we measured the charge on particles of different material compositions suspended in atmospheric conditions for long periods of time. In dry environments, the total charge on particles decayed in over 1 week. The decay timescale decreased to days in humid environments. These results were independent of particle material and charge polarity. However, exposure to UV radiation could both increase or decrease the decay time depending on polarity. Our work suggests that the rate of charge decay on airborne particles is solely determined by ion capture from the air. Furthermore, using a one-dimensional sedimentation model, we predict that atmospheric dust of order 10 $\mu$m will experience the largest change in residence time due to electrostatic forces. 
\end{abstract}

\maketitle

Atmospheric dust is an important component of local and global climate systems. At regional scales, events such as volcanic eruptions, dust storms, and forest fires may rapidly and dramatically alter local atmospheric dust budgets. These acute increases in solid mass loading represent hazards to populations, natural environments, and infrastructure \cite{young2014surface, kaspari2015accelerated}. On a planetary-wide scale, atmospheric dust interacts with short- and long-wave radiation, tuning the Earth's energy balance \cite{tegen1996modeling}. In turn, this modulation has profound impacts on sea and land surface temperatures, atmospheric circulation, and weather \cite{Idso1977,Chaibou2020}. Solid particles may also serve as cloud condensation and ice nuclei, influencing the formation of clouds and impacting precipitation rates \cite{ryder2019coarse}. Airborne particles have the ability to carry nutrients, toxins, and bacteria across long distances. Indeed, Saharan desert dust has been recognized as an important source of phosphorous for the Amazon rain forest \cite{yu2015fertilizing}. Similarly, lofted dust can modify atmospheric chemistry by providing reactive substrates for various chemical species \cite{usher2003reactions}. Beyond Earth, atmospheric dust likely influences surface processes on Venus, Mars, Io, Titan, and Gliese J1214b  \cite{thomas1985dust, Mendez2018KCL, rodriguez2018observational, mcdonald2022io}. 

Since the 1970s, evidence indicates that climate models fail to accurately represent the quantity of coarse-mode particles (with diameters $D>5$ $\mu$m) in the atmosphere \cite{weinzierl2017saharan, adebiyi2020climate} by a factor of 4 \cite{adebiyi2020climate}. Although dust transport models predict that coarse-mode dust deposits out of suspension rapidly, observations consistently find large particles at distances well beyond the maximum ranges estimated numerically.
For instance,  Denjean et al. showed that the effective diameter of coarse-mode Saharan over the Mediterranean remained unchanged for up to a week after being lofted into suspension \cite{denjean2016size}. Likewise, modelling shows that particles in the range of 20–30-$\mu$m should settle out of the Saharan Air Layer in approximately 1.5-3 days. Yet, grains with these diameters have been detected over the Caribbean after 4,000 km and 5 days of transport \cite{weinzierl2017saharan}. 


Recently, a number of investigators have suggested that electrostatic forces may significantly influence the residence time of atmospheric dust. As particles are injected into the atmosphere by aeolian action, splashing, chemical processes, and comminution they may become charged. \cite{kok2008electrostatics}. Charging mechanisms include fracto- and triboelectric charging \cite{james2000volcanic, mendez2021charge,Waitukaitis2014}, radioactive decay (specifically, radium and thorium) \cite{Aplin2014}, gas ionization \cite{okuzumi2009electric}, and the so-called inductive mechanism \cite{pahtz2010particle}. Surprisingly, a number of investigations have demonstrated that airborne particles may remain charged even at great distances from where they emanate. For example, anomalously high volumetric charge densities were found in an ash cloud 1,200 km from its source at the Eyjafjallaj\"okull volcano \cite{harrison2010self}, and Saharan dust over Scotland can carry an edge charge density several times larger than that of typical stratiform clouds \cite{harrison2018saharan}. 

\begin{figure*}[h]
\centering
\includegraphics[width=6.5in]{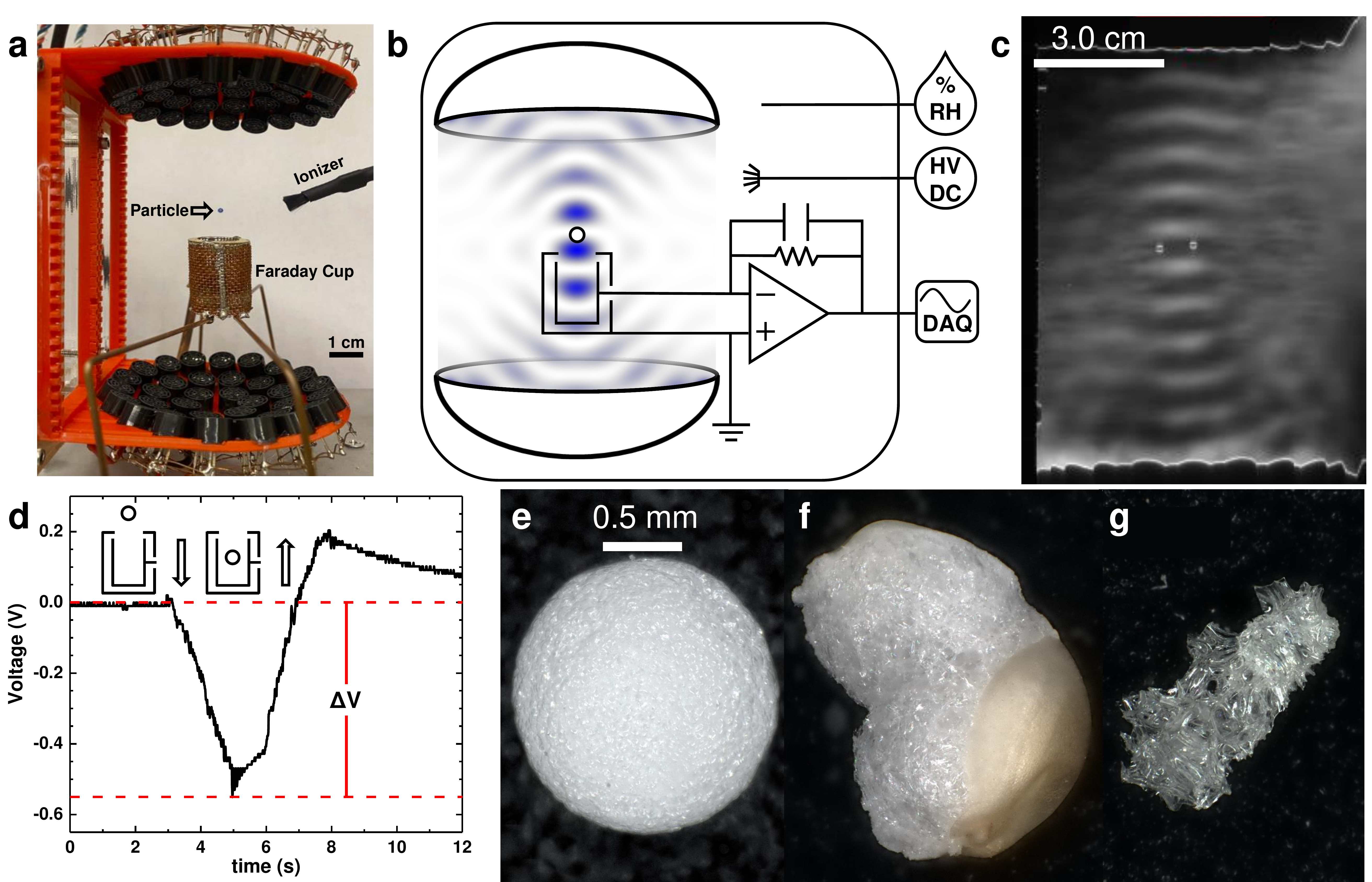}
\caption{(a) A TinyLev acoutsic levitator suspends a 2 mm polystyrene bead above the acoustically-transparent Farday cup. (b) Diagram of the charge measurement system inside the environmental chamber with a theoretical acoustic field. (c) Schlieren image of the acoustic field with a single levitated particle \cite{SettlesSchlieren,Crockett2018}. A second, virtual image of the particle is visible in the mirror, see \textbf{SI}. (d) Output of the amplification stage as the particle was lowered into and raised out of the Faraday cup. $\Delta$V that corresponds to the charge on the particle. (e) Expanded polystyrene bead (EPS). (f) Toasted amaranth grain. (g) Pumice from the 1932 eruption of the Quizap\'u volcano (Maule, Chile).}
\label{Fig1}
\end{figure*}

Accurately assessing the degree to which electrostatic forces influence the transport of large particles requires better constraints on 1) the magnitude of the electric fields within atmospheric dust layers, 2) the mechanisms by which particles charge before and after being lofted, and 3) the ability of electrified grains to retain charge once airborne. Improvements to our understanding of the meso- to macroscale electrostatic characteristics within dusty environments may come from a combination of airborne and ground-based measurements, complemented by numerical modeling \cite{harrison2010self, mendez2017electrification, adebiyi2020climate, zhang2020reconstructing}. However, elucidating charge evolution on particles fundamentally requires long-term measurements at the grain scale. Such measurements are difficult to do \textit{in-situ} and, until recently, laboratory experiments have been unable to isolate charged grains from other surfaces to mimic airborne transport.

Here, we investigate the longevity of charge on isolated particles of various compositions suspended in air. We implement a non-contact charge measurement technique that can monitor charge decay of a single acoustically-levitated particle for days to weeks. Unlike previous investigations which allowed charge to leak away across contact points, the present experiments characterize charge loss occurring only at the particle-gas interface. Our results suggest that lofted particles can retain charge for weeks, with the decay rate depending only on environmental factors like relative humidity, charge polarity, and irradiation. Relative humidity (RH) decreases the half-life ($t_{1/2}$) of charge decay by roughly a factor of 10 at saturation, while particle composition, size (in the range of 0.5 to 2 mm), and polarity seem to have little effect. Asymmetries in the affect of UV radiation requires further study, but can provide insight into the relative influence of radiation on particles in Earth's atmosphere. We present a simple ion recruitment model that can accurately predict decay times and curves below 60-70\% RH. Together, our experiments and models suggest that electrostatic forces may significantly influence the residence times of particles with diameters between 1 and 100 $\mu$m.

\section*{Results and Discussion}

We measured the decay of surface charge on single particles lofted in air using a TinyLev Acoustic Levitator (TAL) \cite{TinyLev} housed in a environmental chamber. TAL is an open-source, 3D-printed instrument capable of manipulating micron- to millimeter-sized particles in a non-contact manner (see Methods; \textbf{Fig. \ref{Fig1}a} and \textbf{b}). We modified TAL with the ability to measure charges on suspended particles with a resolution of 1 fC. Particles placed in the acoustic trap are initially charged  by ionizing the air in the chamber with a high-voltage supply connected to a bundle of sharp carbon needles. Free ions and electrons rapidly adhere to all surfaces including the isolated particle. Using a Gerdien tube condenser ion counter (AlphaLab, Inc), we estimate the ion density at the location of the levitating particle to be 10-15 $\times$ 10\textsuperscript{6} cm\textsuperscript{-3} during the charging period. The ion concentration rapidly decays a few seconds after the ionizer is shut off, presumably due to collisions with grounded metallic surfaces in the apparatus. Once charged, a particle's surface only discharges across the gas-solid boundary.

TAL can hold a trapped particle static and move it along its primary axis (\textbf{Fig.\ \ref{Fig1}c}; \textbf{Fig.\ S1}; \textbf{Movie S1}). We use this capability to perform non-contact charge measurements across extended time frames: a levitating particle is periodically (every 1-5 min) lowered into and raised out of an ``acoustically transparent'' Faraday cup (ATFC) to ascertain its surface charge. The ATFC is connected to charge-sensitive electrometer which provides an overall sensitivity of 1 pC/V. A typical signal from the output stage of the electrometer during one measurement cycle is shown in \textbf{Fig. \ref{Fig1}d}. The voltage difference $\Delta V$ between the initial baseline and the curve's minimum is proportional to the amount of charge on a particle's surface. Because the particle never touches a surface during the measurement process, charge loss occurs only through interactions with the gas. Additional details about the charge measurement system are included in the Methods.

\begin{figure}[t]
\centering
\includegraphics[width=3.25in]{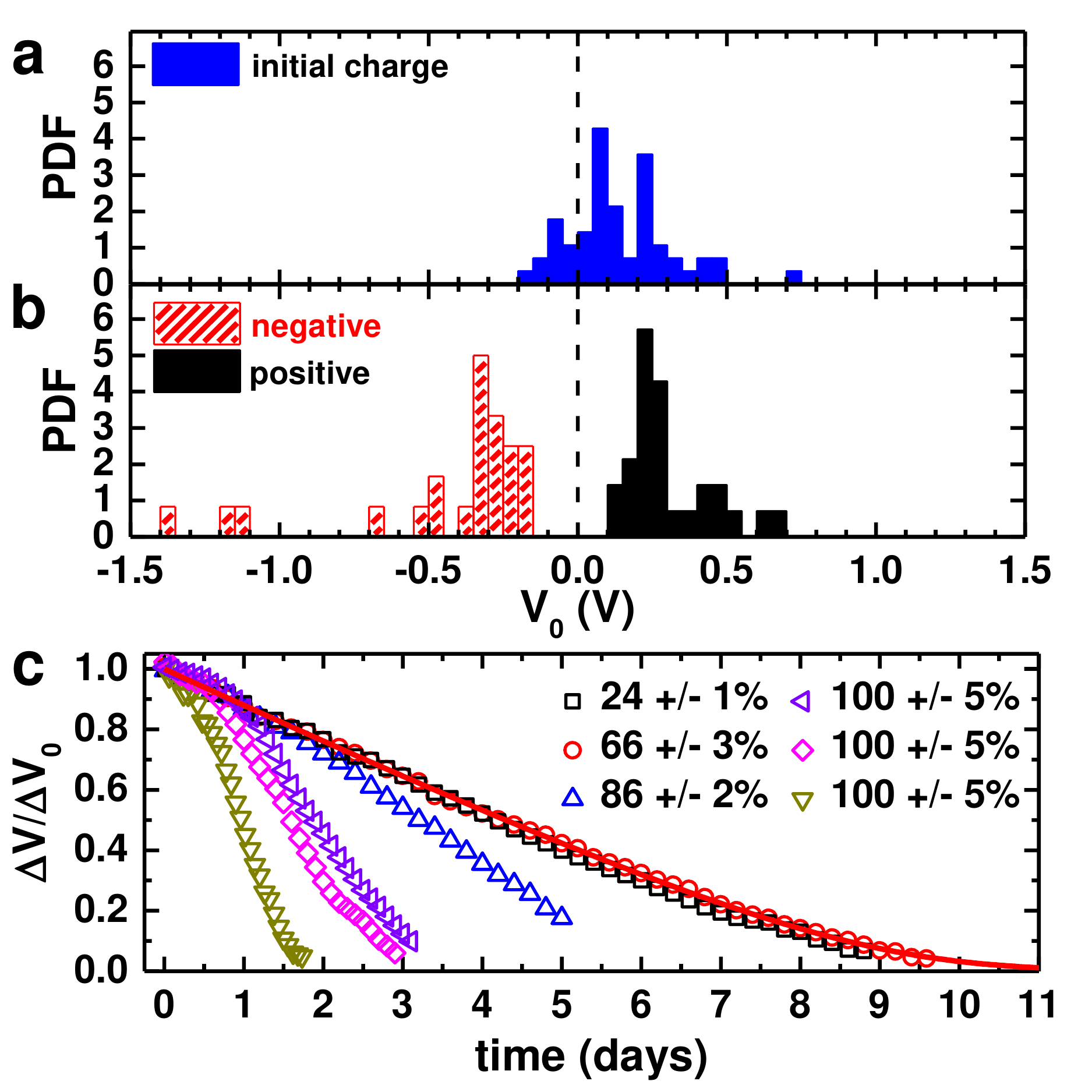}
\caption{(a) Probability density function of charge on EPS particles before ionization. Particles have some initial positive charge after being extracted from container. (b) Distribution of charge on particles immediately after ionization with negative or positive bias. (c) Normalized voltage $\Delta V / \Delta V_0$ vs time. The low RH data is fit well by Eq. \ref{fit} as shown by the red line which fits the 66\% RH decay. The fit parameters are $Q_0$ = 1, $Q_c$ = 0.11, and $K$ = 1.24.}
\label{Fig2}
\end{figure}

To approximate the diversity of particles suspended in Earth's atmosphere (from silicate dust to pollen to microplastics), we explored the decay of charge on particles of numerous compositions: expanded polystyrene (EPS), aerogel, toasted amaranth (Peru), and volcanic pumice (Quizap\'u, Popocat\'epetl, and Kos). Exemplary grains are rendered photographically in \textbf{Fig. \ref{Fig1}e-g}. All particles, despite a wide variation in shape, had spherical-equivalent diameters (SED) ranging between 1-2 mm. EPS particles were transferred to the experimental setup directly from their packaging container (nylon bag). Conversely, aerogel and pumice samples crushed to the appropriate size and were then stored in a desiccator at $<$10\% RH until beginning an experiment. Amaranth samples were toasted immediately prior to inserting them into TAL. We note that particles inserted into TAL may be precharged as a result of contact electrification during handling. For instance, \textbf{Fig. \ref{Fig2}a} shows that EPS particles taken from their plastic packaging bag generally carry positive charge (consistent with the relative positions of polystyrene and nylon on the triboelectric series \cite{zou2019quantifying}). However, the ionization process is capable of erasing this initial bias (\textbf{Fig.\ \ref{Fig2}b}; \textbf{Fig.\ S2}).

The environmental chamber which encloses TAL allowed us to explore the effect of atmospheric water (at 25$^\circ$C) on particle charge decay. The decay dynamics for a subset of EPS particles are rendered in \textbf{Fig. \ref{Fig2}c}. In dry environments, the functional form of the charge decay was linear at early times, giving way to nearly exponential behavior at long times (see black and red curves \textbf{Fig. \ref{Fig2}c}). For RH $<30\%$, all particles experienced charge decay over a characteristic half-life, $t_{1/2}$, of 2-8 days. At higher RH, a more expedient decay rate followed a logistic form (green, pink, and purple curves).  This timescale decreased to 1 day or less for RH approaching 100$\%$. Yet, even the shortest decay times in our experiments greatly exceed those of seconds to hours reported in several previous works for broad ranges of RH \cite{Tada1995Direct,Kumara2011Surface}. We note that those efforts did not use isolated charged surfaces (i.e. tested materials were clamped, tethered, or otherwise secured extraneous surfaces). As such, we suspect that shorter decay times there reflect conduction processes in addition to charge exchange between solid and gas. Indeed, Burgo et al. \cite{Burgo2011Electric} found multi-day decay times for charged, polyethylene slabs in contact with aluminum, but short decay times at higher RH, possibly resulting from the growth of a conducting water layer on the surface.   

Using the TAL system, particles only discharge by exchanging charge carriers with the atmosphere. Previous efforts have invoked ion recruitment processes to account for charge gain or loss on particle and aerosols surfaces \cite{gunn1954,Lacks2022JoE}. The conventional prediction from such models is that particles reach electrostatic equilibrium in minutes to hours, not days. The dramatic departure from theory implied by the long charge decay times observed in our experiments may indicate that the regions near the charged object are depleted of free ions. Such depletion was recently suggested by Heinert et al. \cite{Lacks2022JoE} to explain the multi-day decay of charge on a magnetically levitated conducting disk.

Particles in our experiments have radii of $a\approx$ 1 mm and initial charges $|Q_0|\lesssim$ 1 pC = $6.25\times10^6$ elementary charges. These conditions correspond to approximately 0.5 charges for every square micron of particle surface area. While this charge density may seem dilute, the background ion density, as measured by the Gerdien tube condenser, is less than 1 ion/mm$^3$ for both positive and negative ions. The dynamics of charge neutralization in our experiments can be understood by considering the ratio of the potential energy to thermal kinetic energy for a single airborne ion: $e\phi/k_BT=eQ/4\pi\epsilon_0 r k_B T$. In this framework, $k_B$ is Boltzmann's constant, $T$ is the temperature, $\epsilon_0$ is the permittivity of free space, $e$ is the elementary charge, and $r$ is the separation between the ion and the charged particle. We make the simplifying assumption that oppositely-charged ions near the particle are always captured, whereas ions far from the particle can escape capture. The boundary between these regimes is estimated as $|e\phi/k_BT|\approx 1$. 

For the typical parameters described above when $|Q_0|$ = 1 pC, this boundary corresponds to a radius $r\approx$ 35 cm, which is larger than the size of our experimental chamber. Thus, in the early stages of charge decay, we can assume that \emph{every} oppositely charged ion generated in the chamber is captured by the charged, millimetric particle. Indeed, the near linear decays seen in \textbf{Fig.\ \ref{Fig2}c} for low RH suggest that discharge rates are likely determined by and equal to the rate of ion production in air surrounding the particle. However, as the particle loses surfaces charge, the capture radius becomes significantly smaller than the size of the experimental chamber. For example, when $Q=0.1Q_0$, $r\approx$ 3.5 cm, and the decay rate of charge will depend on $Q$. Moreover, free ions may be funneled away by the flow system that maintains the chamber at low RH or may be neutralized by ions of opposite polarity before being captured by the particle.

A very simple model for the charge decay that includes both of these regimes can be written as:
\begin{equation}
\dfrac{dQ}{dt}=-\dfrac{K}{1+Q/Q_c}Q,
\end{equation}
where $K$ is a characteristic rate, and $Q_c$ is a characteristic charge representing a crossover between these regimes. A similar equation results from the capture of molecules by diffusion to a spherical particle covered in absorbing patches, as first discussed by Berg and Purcell in the context of chemoreception \cite{berg1977}. This equation can be readily solved for $Q(t)$:
\begin{equation}
Q(t)=Q_c W\left[\dfrac{Q_0\exp{(Q_0/Q_c-Kt)}}{Q_c}\right],
\label{fit}
\end{equation}
where $W$ represents the Lambert function, and $Q_0$ is the initial charge on the particle. This 3-parameter function shows excellent agreement with the data in dry conditions, as shown by the solid line in \textbf{Fig.\ \ref{Fig2}c}. It is important to note that Eq.\ \ref{fit} does not depend on the size of the particle, which may explain similar decay times observed in experiments with much larger, centimeter-scale objects \cite{Lacks2022JoE,Burgo2011Electric}. 

Even though data for RH as high as 60-70\% follows the same trend as for very dry conditions, real lofted particles are often found in water-rich environments near saturation. As \textbf{Fig.\ \ref{Fig2}c} shows, the decay rate of charge increases significantly with higher RH. Generally, equilibrium ion concentrations increase with RH \cite{Carlon1981}, but the rate at which a system recovers to its equilibrium concentration is not well-known. In our experiments, we assume the charged particle and surfaces of the TAL apparatus quickly deplete nearly all ions in the chamber once the ionizer is shut off. Thus, we expect that higher RH acts to return the ion concentration to equilibrium more rapidly. Initially, the charge decay rate is small before increasing at later times as the ion concentration returns to equilibrium, resulting in a logistic-shaped decay curve, as seen in \textbf{Fig.\ \ref{Fig2}c}. This particular decay, characterized by an \emph{increase} in decay rate during the lifetime of the experiment, has been observed in other studies \cite{Lacks2022JoE,Burgo2011Electric}. It is unclear if this behavior is solely related to RH, or other uncontrolled conditions.

\begin{figure}[t]
\centering
\includegraphics[width=3.25in]{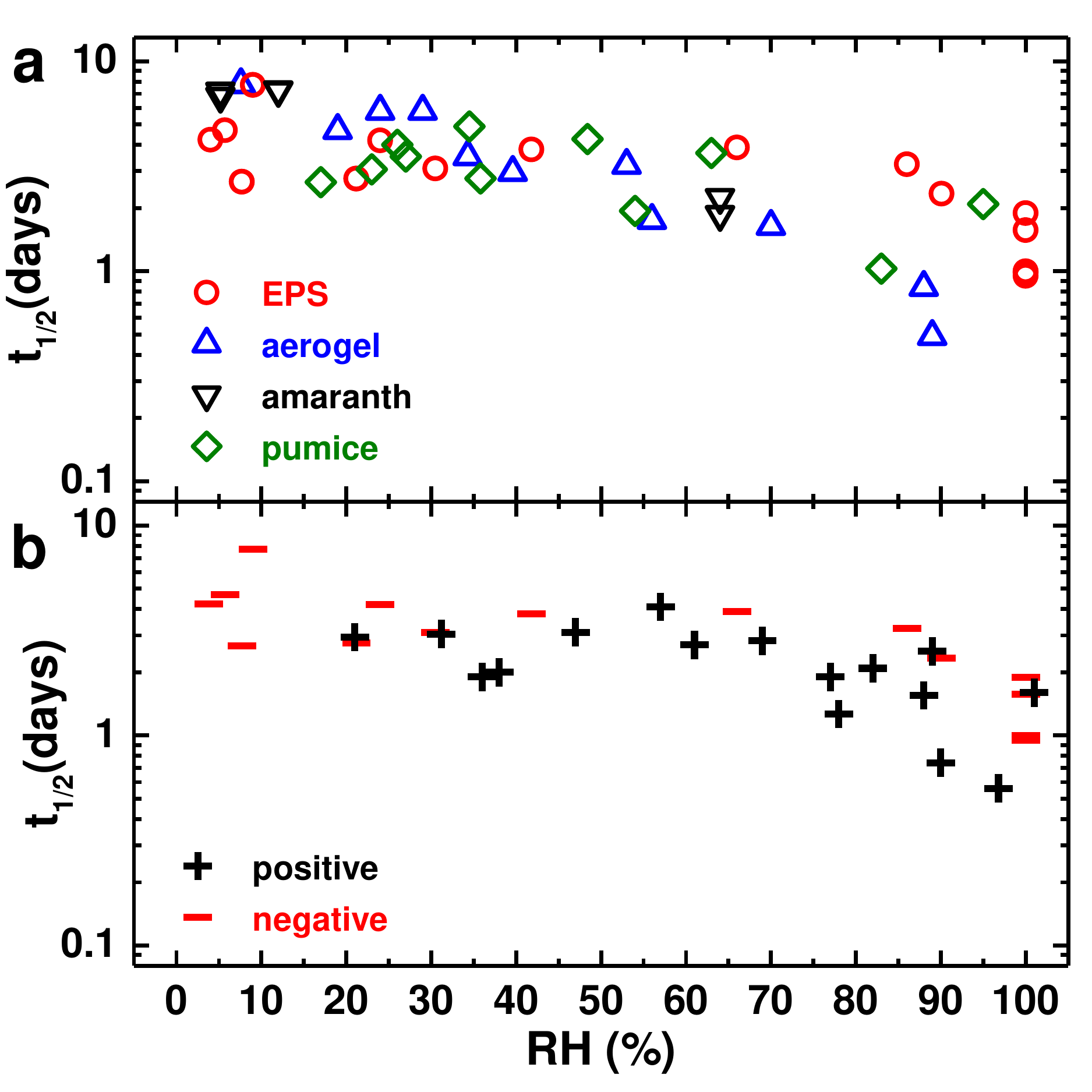}
\caption{$t_{1/2}$ vs RH for (a) different materials that were negatively charged, and (b) EPS that was positively or negatively charged. We observed similar trends for $t_{1/2}$ with increasing RH for all materials used in (a). In (b) we observe a slightly smaller half life at lower RH for EPS particles, possibly due to differences in positive and negative ion concentrations.}
\label{Fig3}
\end{figure}

Although data at higher RH cannot be fitted with Eq.\ \ref{fit}, we can compare the half-life, $t_{1/2}$, of charge decay across all materials and RH. Remarkably, \textbf{Fig.\ \ref{Fig3}a} shows no observable variation of $t_{1/2}$ for the different particles used in our experiments. These particles vary in their shape, size, porosity, and hydrophobicity. Additionally, $t_{1/2}$ decreases with RH by roughly a factor of 10 at saturation for all particle types. This is consistent with our model since the capture probability of a given ion is only dependent on the atmospheric conditions and the total charge on the levitating particle. It is possible that in ion rich environments the charge decay of these particles would vary based on local surface properties or combinations of ion transport mechanisms such as diffusion, electrostatic drift, and convection. In our experiments, the acoustic levitator generates a small amount of heat, which nonetheless causes convection that is visible in the Schlieren imaging of the acoustic field (Movie S1). However, these environmental variations occur on much shorter timescales where we observe no large fluctuations in the rate of charge decay. 

\begin{figure}[t]
\centering
\includegraphics[width=3.4in]{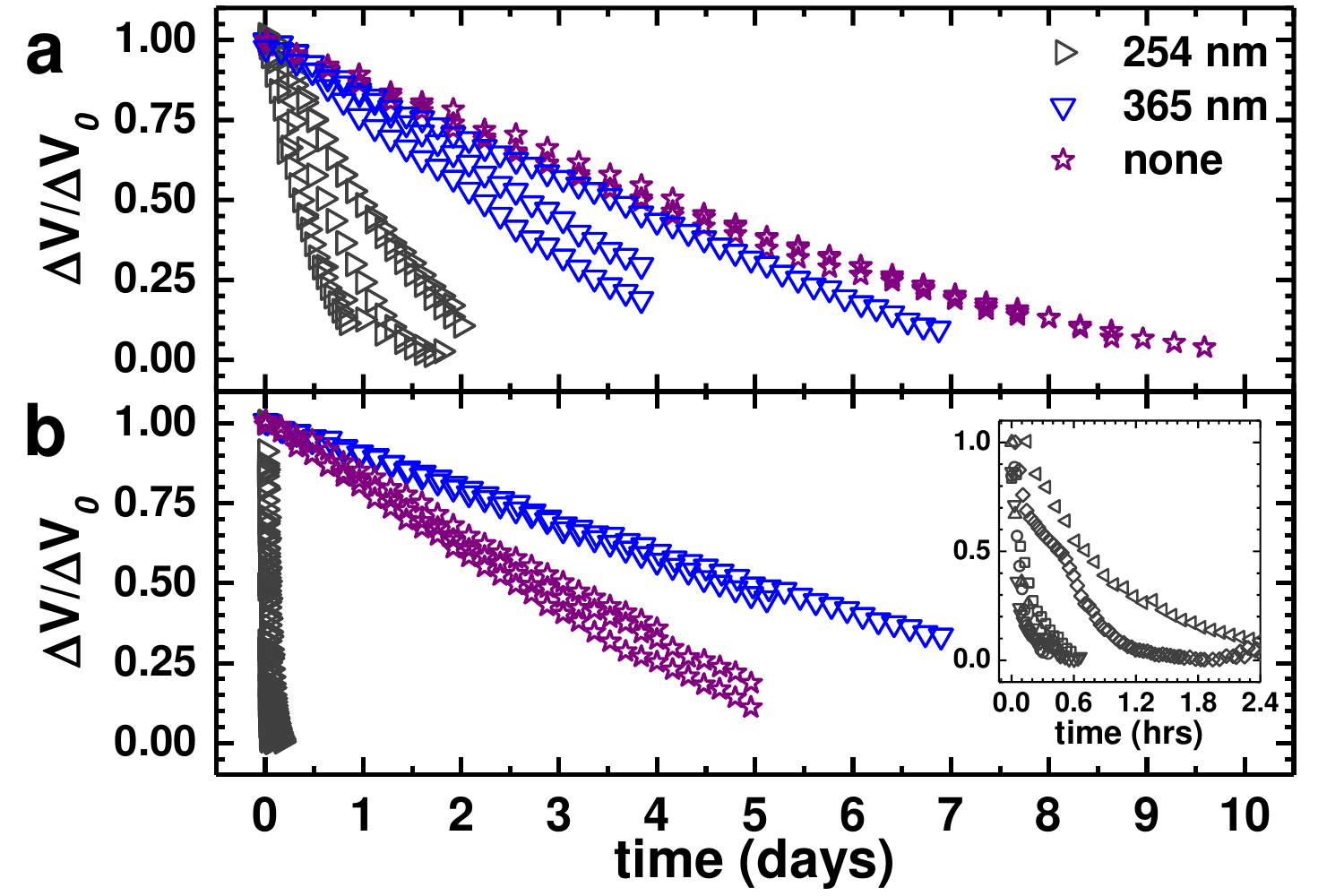}
\caption{$\Delta V / \Delta V_0$ vs. time for particles irradiated by UV of different wavelengths, as indicated in the legend. Panel (a) represents negatively-charged EPS, and panel (b) represents positively-charged EPS. The inset of (b) shows a zoom-in of positively-charged particle data with the shortest decay times. The different symbols in the inset correspond to repeated experiments.}
\label{Fig4}
\end{figure}

Atmospheric dust may carry net positive or negative charge \cite{mendez2021charge, james2000volcanic, jungmann2022aggregation}. Figure \ref{Fig3}b shows results for negatively and positively charged EPS particles, indicating little to no difference in $t_{1/2}$. If discharge is mostly determined by ion recruitment from the atmosphere, then the decay rate should represent the relative concentrations of positive and negative ions. Measurements suggest a roughly equal number of positive and negative ions in our experimental chamber, explaining the similarity between decay rates observed from positively and negatively-charged particles. 

Lofted particles in Earth's atmosphere also experience varying amounts of UV radiation, ranging from UVC at the highest altitudes, to UVA near the Earth's surface. To simulate this exposure, in our experiments we used two different types of UV light bulbs. Naively, one may expect that UV radiation always accelerates charge decay as high energy photons neutralize surface charge. Previous studies have shown that the efficiency of UV light on discharging can vary dramatically in a narrow range of wavelength \cite{Ugolini2008}. Strikingly, we observe that UV radiation can even \emph{extend} the life of charge on some particles. Figure \ref{Fig4} shows data for EPS particles with negative (\ref{Fig4}a) and positive (\ref{Fig4}b) initial charge in dry conditions (low RH). With no UV radiation, particles exhibit a multi-day decay timescale ($t_{1/2}$), as expected. For UVA radiation (365 nm, 3.40 eV), $t_{1/2}$ decreased slightly to 2-3 days for negatively-charged particles, yet $t_{1/2}$ \emph{increased} significantly for positively-charged particles. 

This asymmetry can be explained by noting that UV radiation can potentially ionize organic atmospheric impurities that could deplete the local ion concentration, or even by direct photoelectric charging of the particle. The charged surface states can be highly variable given the complex particle materials, and the potential presence of water. Our experiments cannot disentangle these effects at the moment, however, the asymmetry is highly relevant for most airborne particles since UVA radiation can penetrate to Earth's surface. For UVC radiation (254 nm, 4.88 eV), which is absorbed in Earth's upper atmosphere, particles of either charge polarity experienced a decrease in $t_{1/2}$, yet positively charged particles decayed in less than 1 hour. This drastic decrease in the decay rate is almost certainly due to photoelectric electrons emitted from the surface of the copper mesh comprising the Faraday cup. The work function of copper is 4.7 eV.

How does the multi-day decay of charge affect transport in the atmosphere? The dynamics of a particle with a spherical equivalent diameter (SED) of $2a$ settling out of the atmosphere are governed by gravitational, $F_g$, drag, $F_d$, and electrostatic forces, $F_e$. Here, we assume the simplest 1D model of this process (see SI for full treatment):
\begin{equation}
    m\dot{v} = F_g + F_d + F_e,
\end{equation}
\noindent 
where $m$ is the particle mass and $\dot{v}$ is the acceleration in the direction of gravity. We can quantify the impact of charge decay on atmospheric residence time by computing the difference sedimentation time $t_{set}$ between a charged particle and an otherwise identical neutral particle. For this analysis, we consider particles with SEDs in the range of 1 - 100 $\mu$m and densities between 1000 - 2900 kg/m\textsuperscript{3}. Earth's fair weather electric field is on the order of 0.1 kV/m and points toward the surface. However, $\sim10$ kV/m is typical in dust storms \cite{zhang2020reconstructing}, and $\sim100$ kV/m has been measured during foul weather \cite{dwyer2014physics}. As such, we employ a conservative electric field range spanning $\pm$5 kV/m. Furthermore, we assume that particles are initially charged to the theoretical maximum limit of $\sigma\sim 10^{-5}$ C/m\textsuperscript{2} and then decay exponentially with a half life of 4 days. Lastly, we determined settling times for particles falling from an altitude of 5 km, an elevation at which dust has been detected in the Saharan Air Layer \cite{nicoll2010observations}. The model implements a simple atmospheric profile to account for changing pressure as a particle descends. We note that our model excludes the effects of turbulence and convective uplift.

\begin{figure}[t]
\centering
\includegraphics[width=3.45in]{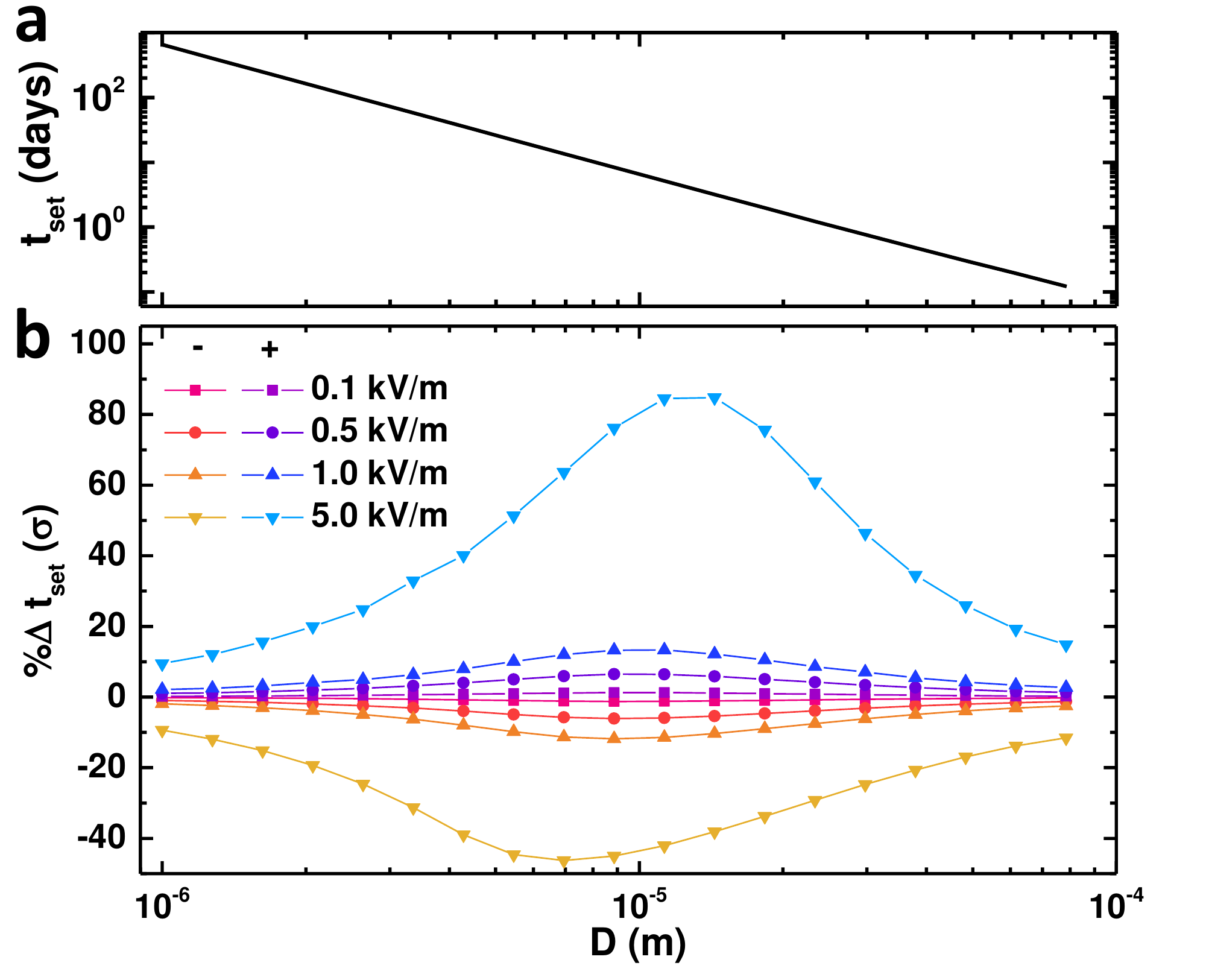}
\caption{Model results for silicate particle sedimentation times assuming no turbulent mixing from an altitude of 5 km. (a) Without electrostatic forces, $t_{set}$ is determined by gravitational and drag forces, ranging from years to hours for particle diameters 1-100 $\mu$m. (b) Percent change in settling time for a positively-charged particle in an external electric field with a surface charge density $\sigma=10^{-5}$ C/m$^{2}$ that decay exponentially with $t_{1/2}$ = 4 days. The symbols represent different ambient electric fields ($+$, upward pointing; $-$ downward pointing).}
\label{Fig5}
\end{figure}

Settling times from an altitude of 5 km can range between a few hours for the largest particles to >10\textsuperscript{2} days for the smallest grains (see \textbf{Fig. \ref{Fig5}a} and \textbf{SI}). Because small, neutral particles take months or even years to settle, electrostatic effects with half-lives of a few days should not affect $t_{set}$. Likewise, electrostatic forces play a minimal role for large particles since these are dwarfed by inertial forces. Conversely, our numerical experiments show that when $t_{set}\approx t_{1/2}$, electrostatic effects can substantially modify particle residence times. Indeed, we find that charge most strongly influences the dynamics of 2900 kg/m$^3$ particles (mineral dust) with SEDs of 5-20 $\mu$m by increasing $t_{set}$ by 80\% (\textbf{Fig. \ref{Fig5}b}). For lighter 1000 kg/m$^3$ particles (e.g. microplastics), the effects electrostatic forces are shifted toward particles with SEDs of 10-30 $\mu$m and can modify the settling times by up to 175$\%$ (\textbf{Fig.\ S3}).

Beyond more complex fluid dynamics, we note that the above model neglects additional electrostatic processes that may retard, reverse, or accelerate charge loss. As aforementioned, charged dust plumes have been detected at large distances from their sources, suggesting that in-situ charging mechanisms keep lofted particles electrified well beyond the timescales measured here \cite{nicoll2010observations, harrison2010self}. Additionally, we suspect that other environmental factors (such as pressure and temperature) may influence the retention of charge, and these should be explored in future work. However, our first-order analysis highlights the effects electrostatic forces may have on the transport of particles with $2a \sim 10$ $\mu$m. Whether these forces help retain particles in the atmosphere (or more easily remove them) requires better understanding of the charge distribution in dust layers and electric fields at elevation. 

\section*{Conclusions}

Using acoustic levitation and non-contact charge measurement, we find that isolated particles of any material can retain charge for weeks. In general, RH decreased the half-life of charge ($t_{1/2}$), but the effect was less pronounced than reported in previous works \cite{Tada1995Direct,Burgo2011Electric}. Additionally, $t_{1/2}$ was insensitive to the charge polarity of the particles for a broad range of RH. However, exposure to UV light either accelerated or arrested charge decay, depending on charge polarity and UV wavelength. Both RH and irradiation control the local concentration of ions in the air surrounding a charge particle, and ultimately determines the longevity of charge on a particle's surface. Using our experimental data as input to a simple 1D model, we find that electrostatic forces significantly modulate the residence times of airborne coarse mode dust with diameters 5-30 $\mu$m. Our experiments provide the first measurements of charge decay on lofted particles, and demonstrate that electrostatic forces and charge decay must be considered in aeolian dust transport models. 

\section{Math Methods}
\subsection*{Environment set up}
The RH of the chamber was varied by combining boiling water and dry building air. RH values were maintained within 10$\%$ variance over the entire experiment, with the majority of data points within 5$\%$ RH variance. For data presented in \textbf{Fig.\ \ref{Fig3}}, the environment was set before charging the particle. For UV experiments, the UV bulbs were suspended at the inside wall of the chamber, horizontally aligned with the particle. The chamber was left partially open to laboratory air, and the initial $\Delta V_0$ was measured before UV irradiation. The 254 nm bulb produced irradiance at the particle's position of 1.06$\times10^{-1}$ W/m$^2$. The 365 nm bulb had irradiance of 7.07$\times10^{-3}$ W/m$^2$ at the particle's position.
\subsection*{Acoustic levitation}
The acoustic levitator follows the open-source ``TinyLev" design presented in  Marzo \textit{et al.} \cite{TinyLev}. TinyLev consists of two semi-spherical arrays of 40 kHz transducers separated by a distance 11.5 cm. Generally, both hemispheres are driven in phase, creating a standing wave which hold a particle static. However, by slightly changing the phase of one hemisphere, a particle in the acoustic trap may be moved up or down. We utilize this technique to lower and raise the particle in and out of the Faraday cage in a non-contact manner.  Acoustic levitation can support any material smaller than the half-wavelength of sound ($\lambda/2$), and below the mass density limit, set by the power of the transducers. 
\subsection*{Charge Measurement}
Levitated particles are charged for 10 s with a corona ionizer attached to a Bertan Series 225 high-voltage power supply at $\pm$ 8-10 kV to eliminate any initial charge bias (\textbf{Fig.\ \ref{Fig2}a} and \textbf{Fig.\ S2}). After charging, a 5 min wait period allows charge to dissipate from surfaces in the chamber. We measured the particle charge by acoustically lowering it by 2$\lambda$ into the ATFC, then raising it back to its starting position. This was done slow enough to maintain the integrity of the acoustic trap. The ATFC is connected to an electrometer consisting of a charge-sensitive preamplifier and an inverting stage with a gain of 100. Thus, the electrometer generates voltage pulses that are proportional to the charge on a particle entering the ATFC according to:
\begin{equation}
    \Delta V = -100 \times \frac{Q}{C} e^{-t/RC}.
\end{equation}
\noindent
Here, the time constant of the charge amplifier's feedback loop was $RC$ = 5 s, where $C$ = 1 nF and $R$ = 5 G$\Omega$. 

For most experiments, we performed charge measurements every 1-5 minutes. The rate was increased to 2 measurements per minute for experiments in which particle charge decayed rapidly (high humidity or UV). This process was repeated indefinitely until the end of the experiment.

\begin{acknowledgments} 
This material is based upon work supported by the National Science Foundation under Grant No. 2010524.. 
\end{acknowledgments} 

\bibliography{LevChargeArxiv.bib}

\end{document}